\documentclass[%
 twocolumn,
 amsmath, amssymb,
 aps,
 pre,
 longbibliography
]{revtex4-1}

\usepackage{graphicx}
\usepackage{dcolumn}
\usepackage{bm}
\usepackage[english]{babel}
\usepackage[pdfpagelabels,plainpages=false,bookmarks=true,colorlinks,linkcolor=red,urlcolor=blue,citecolor=blue]{hyperref}

\usepackage{multirow}
\usepackage{diagbox}
\renewcommand{\vec}[1]{{\bf #1}}
\newcommand{\braket}[1]{\langle #1  \rangle}
\newcommand{\ket}[1]{| #1  \rangle}

\renewcommand{\min}{\text{min}}

\newcommand{\Potts}{\text{Potts}}
\newcommand{\clock}{\text{clock}}
\newcommand{\Log}{\text{Log}}
\newcommand{\abs}[1]{|#1|}

\newcommand{\site}[1]{{[#1]}}
\newcommand{\tmp}{\text{tmp}}
\newcommand{\Tr}{\text{Tr}}
\newcommand{\qbraket}[1]{[#1]}
\newcommand{\T}{\vec{T}}
\newcommand{\product}{\text{prod}}

\begin{document}
\title{Non-equilibrium Renormalization Group Fixed-Points of the Quantum Clock Chain and the Quantum Potts chain}

\author{Yantao Wu}

\affiliation{
The Department of Physics, Princeton University
}
\date{\today}
\begin{abstract}
We derive an exact renormalization group recursion relation for the Loschmidt amplitude of the quantum $Q$-state clock model and the quantum $Q$-state Potts model in one dimension. 
The renormalization group flow is discussed in detail. 
The fixed-points of the renormalization group flow are found to be complex in general. 
These fixed-points control the dynamical phases of the two models, giving rise to non-analyticities in its Loschmidt rate function, for both the pure and the disordered system.  
For the quench protocols studied, dynamical quantum phase transitions are found to occur in the clock model for all $Q$s considered, while in the Potts model, they only occur when $Q < 4$. 
\end{abstract}

\pacs{Valid PACS appear here}
\maketitle
\section{Introduction}
In recent years, there has been a surge of interest in the critical phenomena identified in the post-quench out-of-equilibrium dynamics of a quantum system \cite{DQPT_disentangle, DQPT_XY, DQPT_XXZ, DQPT_longrange,DQPT_Potts, DQPT_Fisherzero, DQPT_concurrence,DQPT_nonintegrable,DQPT_transfer,DQPT_nnn_Ising,DQPT_firstorder,DQPT_nonintegrable_dutta,DQPT_merging,DQPT_slowquench,DQPT_semiclassical,DQPT_extendedising,DQPT_domainwall}, known as the dynamical quantum phase transition (DQPT) \cite{DQPT, Heyl_2018, Zvyagin_2016, Heyl_2019}. 
The Loschmidt amplitude $G(t)$ has emerged as a fundamental  quantity in DQPT:  
\begin{equation}
  G(t) = \braket{\psi(t)|\psi_0} = \braket{\psi_0 |e^{-iH_1t}|\psi_0}
  \label{eq:Loschmidt}
\end{equation}
where $\ket{\psi_0}$ is typically the ground state of a pre-quenched Hamiltonian, $H_0$. $\ket{\psi(t)}$ is the quantum state evolving under the post-quenched Hamiltonian $H_1$ for time $t$.  
When $\ket{\psi_0}$ is not an eigenstate of $H_1$, $G(t)$ measures the return probability of the system due to a sudden change in the Hamiltonian. 
$G(t)$ scales with the system size, $L$, such that the following rate function is intensive in the thermodynamic limit \cite{DQPT}:  
\begin{equation}
  l(t) = -\frac{1}{L} \log|G(t)|^2 = -\frac{2}{L} \Re\{\Log G(t)\} 
\end{equation}
where $\Log$ is the principal complex logarithmic function. 
It was first found in \cite{DQPT} that $l(t)$ of the transverse field Ising chain (TFIC) exhibits singular dependence on time in the thermodynamic limit.   
Later on, the universality, scaling, and robustness of the DQPT in the TFIC was explained by a renormalization group (RG) \cite{Wilson_rg,Cardy} calculation \cite{DQPT_RG} on the system Hamiltonian.
So far, no other examples have been treated with an RG analysis.
The two RG fixed-points found in \cite{DQPT_RG} are the stable infinite-temperature fixed-point and the unstable zero-temperature fixed-point of the classical Ising chain.    
It is thus not clear whether genuine non-equilibrium fixed points appear in a general setting \cite{Heyl_2018}.   

In this paper, we generalize the RG procedure in \cite{DQPT_RG} to the transfer matrices of the Loschmidt amplitude, which avoids the mathematical complication of the complex logarithmic function.    
As a result, the fixed-point structure of the RG procedure becomes clearer.  
For example, we will discover a non-equilibrium fixed-point that went unnoticed in \cite{DQPT_RG}. 
As we will show, the non-equilibrium RG fixed-points determine the singularities in the Loschmidt rate function of both the pure and the disordered system.    
In the cases that we will consider for this paper, the singularities determined by the RG analysis take the form of linear-cusps, in consistency with the generic crossing of the leading eigenvalues of the transfer matrix.     
Our emphasis here will be to explain the RG procedure in detail and provide the RG origin of these singularities.   
The RG procedure, however, can be carried out in more sophisticated cases where the critical exponent in the rate function differs from one.
We present this case elsewhere \cite{DQPT_Critical}.

It will turn out that in general the RG fixed-points form a continuous line, indicating the presence of marginal scaling operators (explained in Sec. \ref{subsec:Q=2}).  
Special cases, however, can be constructed for which the RG fixed-points are isolated. 
Because these special cases describe the same universality class as the line of fixed-points, we will study them instead. 
In particular, we study the quench protocol of the quantum clock model and the quantum Potts model where the transverse field is infinite in $H_0$ and zero in $H_1$.  
In the clock model, we will discover that DQPTs occur for all the $Q$s considered, i.e. $Q = 2, 3, 4, 5$ and $6$. 
In the Potts model, however, DQPT will only occur for $Q < 4$.  

The paper is organized as follows. 
In section \ref{sec:RG}, we present the RG procedure, using the clock model as an example. 
In section \ref{sec:pure}, we present the results for the pure clock model. 
In section \ref{sec:pure_Potts}, we give the results of the pure Potts model, using the RG procedure introduced in section \ref{sec:RG}.  
In section \ref{sec:disorder}, the disordered clock model is solved with the knowledge of the RG fixed points found in \ref{sec:pure}. 
In section \ref{sec:conclude}, we discuss and conclude. 
\section{The renormalization group procedure}
\label{sec:RG}
Consider first the $Q$-state clock model of $L$ sites in one dimension with periodic boundary condition with the Hamiltonian \cite{DQPT_Potts}, 
\begin{equation}
  H_{\clock} = -\sum_{i=1}^L J_i(\sigma_i^\dag \sigma_{i+1} + \sigma_{i+1}^\dag \sigma_i) - f\sum_{i=1}^L(\tau_i^\dag + \tau_{i})
  \label{eq:clock}
\end{equation}
where the operators $\sigma_i$ and $\tau_i$ act on the $Q$ states of the local Hilbert space at site $i$, which we label by $\ket{0}_i,...,\ket{m}_i,...\ket{Q-1}_i$.  
In this local basis, the $\sigma_i$ is a diagonal matrix with diagonal elements $\omega^m$ where $\omega = e^{i2\pi/Q}$ and $m = 0, \cdots, Q-1$. 
$\tau_i$ permutes $\ket{0}_i\rightarrow \ket{Q-1}_i, \ket{1}_i \rightarrow \ket{0}_i, \cdots, \ket{Q-1}_i \rightarrow \ket{Q-2}_i$, and together with $\tau^\dag$ acts as a transverse-field.   
Note that while the Hamiltonian in Eq. \ref{eq:clock} is called the Potts model in \cite{DQPT_Potts}, it should be called the clock model, because it is the Hamiltonian limit \cite{Hamiltonian_Limit} of the classical clock model \cite{Clock}. 
The Hamiltonian limit of the classical Potts model \cite{Potts} is given in Eq. \ref{eq:potts} and will also be studied later. 

For the Loschmidt amplitude, in order for the RG equation to be exactly solvable, following \cite{DQPT_RG}, we take the paramagnetic direct product state as the initial state: 
\begin{equation}
\ket{\psi_0} = \ket{\psi}_{\product} = \bigotimes_{i=1}^L \frac{1}{\sqrt{Q}}(\ket{0}_i + \ket{1}_i + ... + \ket{Q-1}_i), 
\end{equation}
and the ferromagnetic Hamiltonian as the evolving Hamiltonian: 
\begin{equation}
H = -\sum_{i}J_i (\sigma_i^\dag \sigma_{i+1} + \sigma_{i+1}^\dag \sigma_i). 
\end{equation}
In this case, $G(t)$ becomes formally identical to a classical partition function \cite{DQPT_RG}:  
\begin{equation}
  \begin{split}
    G(t) &= \sum_{\vec m}  \T_{m_1m_{2}}^{\site{1}} \T_{m_{2}m_{3}}^{\site{2}}\cdots  
  = \Tr(\T^\site{1}\T^\site{2}\cdots)
\end{split}
  \label{eq:G}
\end{equation}
where $\vec m = \{m_1, m_2, ..., m_{L}\}$ is the set of degrees of freedom of this partition function and $m_i = 0, 1, ..., Q-1$ takes the value of a spin at site $i$.   
Here $\T^\site{i}_{m_i m_{i+1}}$ is the transfer matrix of the system between sites $i$ and $i+1$ and depends only on the difference between $m_i$ and $m_{i+1}$ modular $Q$, $m \equiv (m_{i+1} - m_{i}) | Q$ \cite{DQPT_RG}. 
That is, 
\begin{equation}
  \T^\site{i}_{m_i m_{i+1}} \equiv E^\site{i}_{m} = \frac{1}{Q} e^{it J_i 2 \cos(\frac{2\pi}{Q}m)}.  
  \label{eq:transfer}
\end{equation}
Anticipating the disordered system, we allow the transfer matrix to depend on the lattice site $i$.   

To analyze $l(t)$, we perform the decimation coarse-graining \cite{Cardy}, i.e. every other spin is summed away while keeping $G(t)$ invariant.   
In equilibrium RG calculations, upon coarse-graining, one typically considers the transformation of Hamiltonians, i.e. the logarithms of transfer matrices, as is also done in \cite{DQPT_RG}.   
Here, however, because of the complex logarithmic function, renormalizing Hamiltonians brings significant complication.  
We will thus directly deal with the transfer matrices.
The decimation coarse-graining is equivalent to multiplying two neighboring transfer matrices into one:  
\begin{equation}
  \begin{split}
    \text{step 1: }& \T'^{\site{i'}}_\tmp = \T^\site{i}\T^\site{i+1} 
  \\
  \text{step 2: }& \T'^{\site{i'}} = \frac{\T'^{\site{i'}}_\text{tmp}}{(\T'^{\site{i'}}_\text{tmp})_{0s}} 
\end{split}
\label{eq:RG}
\end{equation}
where $(\T'^{\site{i'}}_{\text{tmp}})_{0s}$ is the first nonzero $(\T'^{\site{i'}}_{\text{tmp}})_{0m}$, counting $m$ from $0,1,..$ to $Q-1$. 
Step 2 of Eq. \ref{eq:RG} serves to isolate out the overall multiplicative growth of $\T^\site{i}$ and is necessary for the existence of the RG fixed-points for the pure system.   
As one can check, the renormalized transfer matrix $\T'^{\site{i'}}_{m_{i'}m_{i'+1}}$ still only depends on $(m_{i'+1} - m_{i'})|Q$. 
Thus, the $E^\site{i}_m$s form a complete set of coupling constants, and will be used to parametrize the renormalization.  

The Jacobian of the RG transformation in Eq. \ref{eq:RG} will be needed to compute the critical exponent \cite{Cardy}. 
It is given by:   
\begin{equation}
  \frac{\partial E'_m}{\partial E_n} = \frac{2E_{m-n}\sum_{l=0}^{Q-1}E_l E_{s-l} - 2E_{s-n}\sum_{l=0}^{Q-1}E_lE_{m-l}}{(\sum_{l=0}^{Q-1}E_{l}E_{s-l})^2}. 
  \label{eq:Jacob}
\end{equation}

\section{The pure clock model}
\label{sec:pure}
We now present the RG calculation for the pure clock model with $J_i = 1$, for $Q = 2, 3, 4,$ and 5. 
\subsection{$Q$ = 2}
\label{subsec:Q=2}
To find the fixed-points of Eq. \ref{eq:RG}, let $s = 0$, $E_0 = 1$ and $E_1 = x$, and solve the equation $E'_m = E_m$:  
\begin{equation}
  x = \frac{2x}{1 + x^2}.
\end{equation}
There are three solutions: $E_1 = x = \pm{1}$ and 0. 
One can also check that there are no fixed-points with $s = 1$. 
There are thus three fixed-points of Eq. \ref{eq:RG}: $\vec E^*_a = (1, 1)$, $\vec E_b^* = (1, -1)$, and $\vec E_c^* = (1, 0)$.    
$\vec E_a^*$ and $\vec E_c^*$ correspond to the infinite-temperature and zero-temperature fixed-point Hamiltonians found in \cite{DQPT_RG}. 
The logarithm of $\vec E_b^*$ is not real, and is thus a genuine non-equilibrium RG fixed-point.  
It is the missed fixed-point in \cite{DQPT_RG}.  
The leading eigenvalues of the RG Jacobian at $\vec E^*_a, \vec E^*_b$, and $\vec E^*_c$ can then be computed to be respectively 0, 0, and $2$, suggesting they are respectively stable, stable, and unstable fixed-points. 
Indeed, simulating the RG flow according to Eq. \ref{eq:RG} from the initial transfer matrix in Eq. \ref{eq:transfer}, one discovers that the system flows into $\vec E^*_a$ for $t \in (-\frac{\pi}{8}, \frac{\pi}{8})$, and into $\vec E^*_b$ for $t \in (\frac{\pi}{8}, \frac{3\pi}{8})$, and that the RG flow is the same for $t$ and $t + \frac{\pi}{2}$.  
Separating the two stable phases controlled by $\vec E^*_a$ and $\vec E^*_b$ are two critical times $t_{c,1} = \frac{\pi}{8}$ and $t_{c,2} = \frac{3\pi}{8}$ which flow into the unstable fixed-point $\vec E_c^*$.   
The singular behavior of $l(t)$ is controlled by the eigenvalue of the RG Jacobian at $\vec E_c^*$, which is $\lambda = b^y = 2$, where $b = 2$ is the block size of the coarse-graining and $y = 1$.    
This gives the singular behavior of $l(t)$: 
\begin{equation}
  l(\tau) \sim \abs{\tau}^{d/y} = \abs{\tau}, \hspace{5mm} \tau \equiv t - t_c 
\end{equation}
where $d = 1$ is the spatial dimension of the system. 

Here we explain the significance of the symmetry of the clock model, i.e. the fact that $\T_{m_i m_{i+1}}$ depends only on $(m_{i+1} - m_i)|Q$.  
As one can check, the fixed-point equation of Eq. \ref{eq:RG} only imposes one constraint on a generic fixed-point transfer matrix $\T^* = ((1, x^*), (y^*, z^*))^T$: $z^* = x^* y^*$. 
The system symmetry imposes two additional ones: $z^* = 1$ and $x^* = y^*$.  
Thus, for the transfer matrix in Eq. \ref{eq:transfer}, there are a finite number of RG fixed-points, and if not crossing any DQPT, the RG flow from different $t$ will land on the same fixed-point.   
This is also true for $Q > 2$. 
However, in the absence of the system symmetry, there will be a manifold of solutions to the fixed-point equation of Eq. \ref{eq:RG} and the RG fixed-points will in general depend on $t$ \cite{DQPT_Critical}.  

\subsection{$Q=3$}
\begin{figure}[hth]
\centering
\begin{minipage}{.24\textwidth}
  \centering
  \includegraphics[scale=0.25]{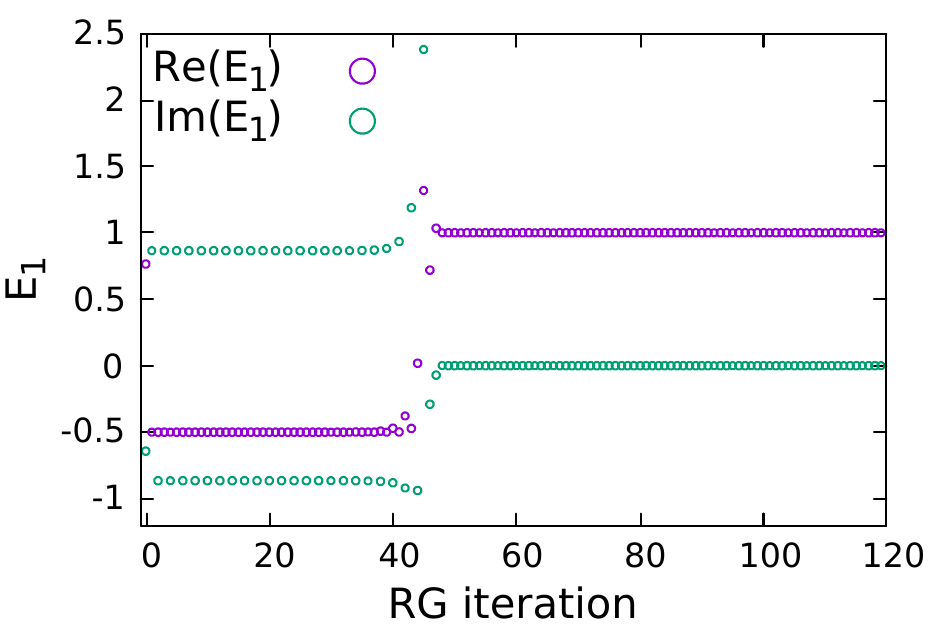}
\end{minipage}%
\begin{minipage}{.24\textwidth}
  \centering
  \includegraphics[scale=0.25]{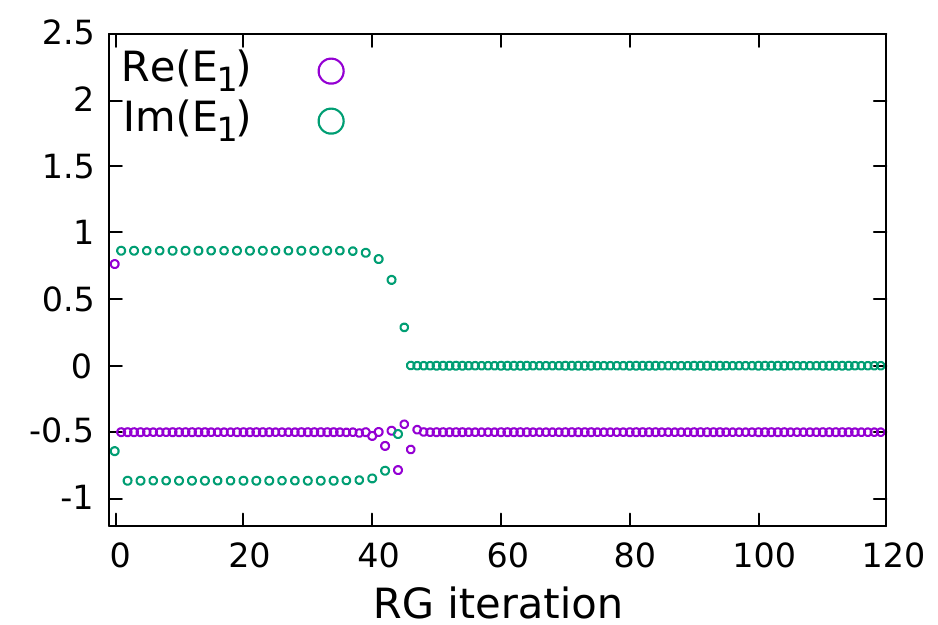}
\end{minipage}
\caption{
The renormalization flow for $Q=3$. The left panel is slightly below $t_c = \frac{2\pi}{9}$, while the right panel is slight above $t_{c}$.
Here we show the real and imaginary parts of $E_1$ during the RG flow. 
}
\label{fig:rg_n3}
\end{figure}

Consider now $Q=3$. 
To look for the fixed-points with $s = 0$, we let $E_0 = 1, E_1 = x_1, E_2 = x_2$ and solve the fixed-point equation of Eq. \ref{eq:RG}:   
\begin{equation}
  x_1 = \frac{2x_1 + x_2^2}{1+2x_1x_2}, \hspace{5mm} x_2 = \frac{x_1^2+2x_2}{1+2x_1x_2}
\end{equation}
This system of equation can be solved by Mathematica, giving seven roots including $x_1 = x_2 = 1$ and $x_1 = x_2 = -\frac{1}{2}$.  
These two solutions correspond respectively to two RG fixed-points, $\vec E_a^* = (1, 1, 1)$ and $\vec E_b^* = (1, -\frac{1}{2}, -\frac{1}{2})$.  
No fixed-points are found with $E_0 = 0$.  
The eigenvalues of the RG Jacobian in the nontrivial eigen-directions at $\vec E_a^*$ and $\vec E_b^*$ are found to be 
\begin{equation}
  \lambda_1 = \lambda_2 = \lambda_3 = 0 \text{ at } \vec E_a^*, \hspace{5mm} \lambda_1 = 2, \lambda_2 = \lambda_3 = 0 \text{ at } \vec E_b^*  
\end{equation}
Simulating the RG flow starting from Eq. \ref{eq:transfer} finds that $\vec E_a^*$ and $\vec E_b^*$ each controls a non-critical phase of $l(t)$.  
Surprisingly, despite the nonzero eigenvalue at $\vec E_b^*$, the system does manage to flow into it for finite periods of $t$. 
In fact, the system flows into $\vec E_a^*$ for $t \in (-\frac{2\pi}{9}, \frac{2\pi}{9})$, and $\vec E_b^*$ for $t \in (\frac{2\pi}{9}, \frac{4\pi}{9})$, and the RG flow is the same for $t$ and $t + \frac{2\pi}{3}$.  
There are thus two critical times $t_{c,1} = \frac{2\pi}{9}$ and $t_{c,2} = \frac{4\pi}{9}$. 
These two critical times, however, do not flow into the other fixed-points found by solving Eq. \ref{eq:RG}.    
For both of them, the system oscillates between a fixed-pair of points: $\vec E_{c,1}^* = (1, \frac{1}{2}(-1 - i\sqrt{3}), \frac{1}{2}(-1 - i\sqrt{3}))$ and $\vec E_{c,2}^* = (1, \frac{1}{2}(-1 + i\sqrt{3}), \frac{1}{2}(-1 + i\sqrt{3}))$, shown in Fig. \ref{fig:rg_n3}. 
The singularity in $l(t)$ is thus not controlled by the fixed-points of the RG transformation in Eq. \ref{eq:RG}, but by the fixed-points of two iterations of Eq. \ref{eq:RG}.     
Multiplying the RG Jacobian computed at $\vec E^*_{c,1}$ and $\vec E^*_{c,2}$ gives the Jacobian of the composed RG transformation: 
\begin{equation}
  \frac{\partial \vec E''}{\partial \vec E} = 
  \begin{pmatrix}
    0 & 0 & 0 \\ 2-2i\sqrt{3} & 4 & 0 \\ 2 - 2 i \sqrt{3} & 0 & 4 
  \end{pmatrix} 
\end{equation}
which has a pair of degenerate eigenvalues $\lambda = 4$.
The block size of the composed coarse-graining, however, is $b' = b^2 = 4$. 
Thus, the critical exponent of $l(t)$ around $t_c$ is still $\frac{d}{y} = \frac{d}{\log_{b'}\lambda}= 1$, giving $l(\tau) \sim \abs{\tau}$. 
The $Q=3$ clock chain has been studied in \cite{DQPT_Potts} using transfer matrix techniques, whose results we agree with exactly. 

The appearance of metastable fixed-points $\vec E^*_b$ is generic and also seen for other $Q$s. 
As $t$ is varied along the real line in the non-critical region, there must be some symmetry of the RG flow which prevents the variation of $t$ from causing any movement along the eigen-direction of the nonzero RG eigenvalue at the metastable fixed-points.      
When $Q = 3$, for example, this symmetry is the equality between the coupling constants $E_1$ and $E_2$.  
Indeed, the equality of the initial $E_1$ and $E_2$ is preserved along the entire RG flow. 
Thus, the direction in the coupling space which is relevant to the quantum dynamics of the clock model is always only along $\delta \vec E = (0, 1, 1)$, orthogonal to the unstable eigen-direction at $\vec E_b^*$, $(0, 1, -1)$. 
\subsection{$Q=4$}
Consider now $Q = 4$.  
The fixed-point equation of Eq. \ref{eq:RG} for $s = 0$ yields 15 fixed-points, including $\vec E^*_a = (1,1,1,1)$, $\vec E^*_b = (1,-1,1,-1)$, and $\vec E^*_c = (1,0,0,0)$.       
The leading eigenvalues at these these points are respectively $0$, $0$, and $2$, suggesting that $\vec E_a^*$ and $\vec E_b^*$ are stable while $\vec E_c^*$ is not.  
The system flows into $\vec E_a^*$ for $t \in (-\frac{\pi}{4}, \frac{\pi}{4})$, and into $\vec E_b^*$ for $t \in (\frac{\pi}{4},\frac{3\pi}{4})$, and the RG flow is the same for $t$ and $t + \pi$.  
There is one critical time $t_c = \frac{-\pi}{4}$ separating the non-critical phases which flows into $\vec E_c^*$.    
The singularity of $l(t)$ is again a linear cusp, suggested by the leading RG eigenvalue $\lambda = 2$ at $\vec E_c^*$.  
\subsection{$Q=5$}
\begin{figure}[hth]
\centering
\begin{minipage}{.24\textwidth}
  \centering
  \includegraphics[scale=0.25]{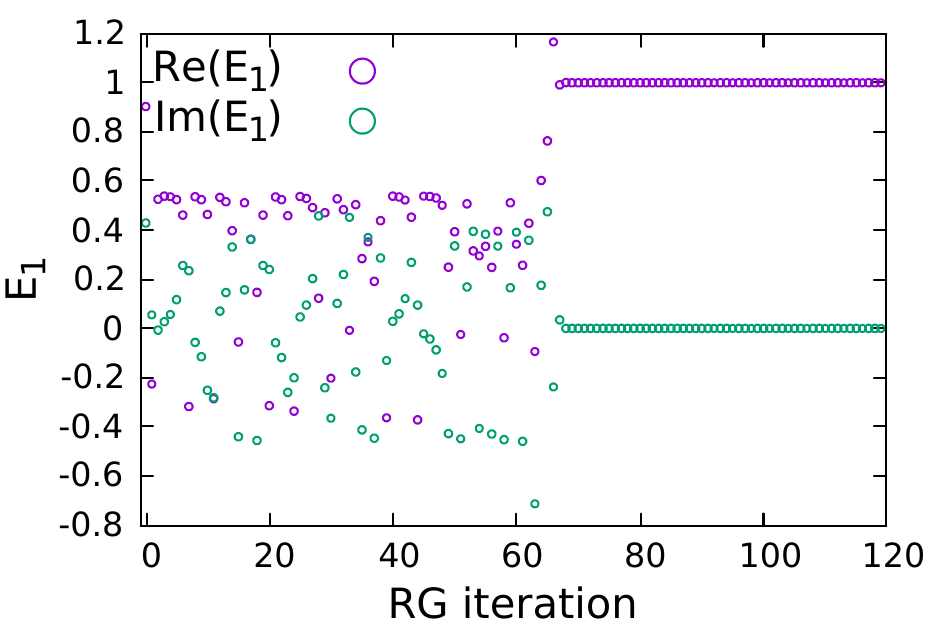}
\end{minipage}%
\begin{minipage}{.24\textwidth}
  \centering
  \includegraphics[scale=0.25]{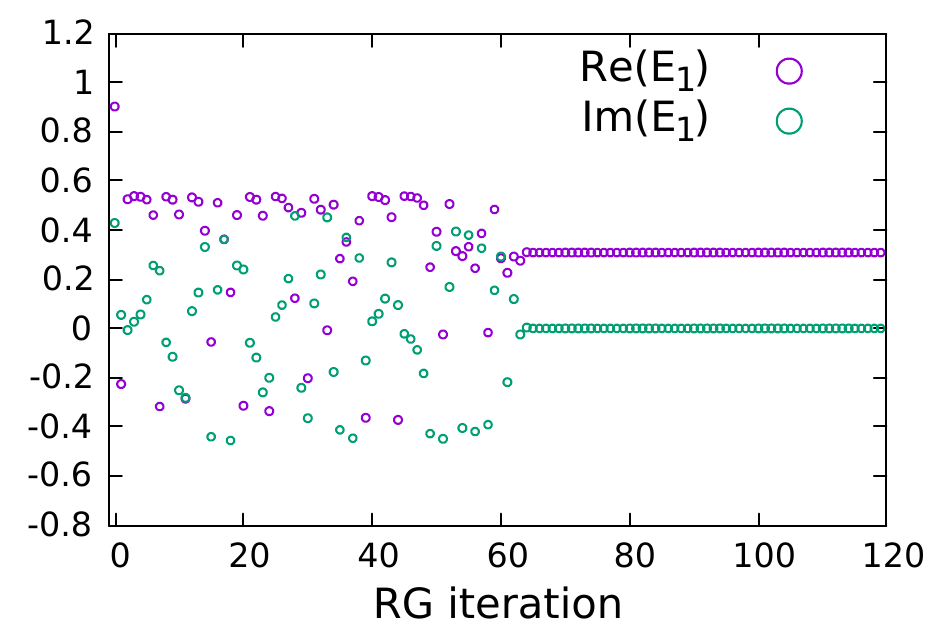}
\end{minipage}
\caption{
The renormalization flow for $Q=5$. The left panel is for $t = 0.7172921525032698574$, slightly below $t_{c,1}$, while the right panel is for $t = 0.7172921525032698575$, slight above $t_{c,1}$.   
Here we show the real and imaginary parts of $E_1$ during the RG flow. 
}
\label{fig:rg_n5}
\end{figure}

Now consider $Q = 5$, which, as we will see, exhibits a chaotic RG flow.  
The fixed-points obtained from solving the fixed-point equation of Eq. \ref{eq:RG} that will interest us are $\vec E_a^* = (1,1,1,1,1)$, $\vec E_b^* = (1, \frac{1}{4}(-1 + \sqrt{5}), \frac{1}{4}(-1-\sqrt{5}), \frac{1}{4}(-1-\sqrt{5}), \frac{1}{4}(-1+\sqrt{5}))$, and $\vec E_c^* = (1, \frac{1}{4}(-1-\sqrt{5}), \frac{1}{4}(-1+\sqrt{5}), \frac{1}{4}(-1+\sqrt{5}), \frac{1}{4}(-1-\sqrt{5}))$.     
Here $\vec E_a^*$ is stable while both $\vec E_b^*$ and $\vec E_c^*$ are metastable, as suggested by the spectrum of the RG Jacobian: 
all of the RG eigenvalues at $\vec E_a^*$ are zero, while both $\vec E_b^*$ and $\vec E_c^*$ have one eigenvalue equal to 2 and four zero eigenvalues. 
In fact, the system flows into $\vec E_a^*$ for $t \in [0, t_{c,1})$, $\vec E_b^*$ for $t \in (t_{c,1}, t_{c,2})$, and $\vec E_c^*$ for $t \in (t_{c,2}, t_{c,3})$, and appears to repeatedly revisit $\vec E_a^*$, $\vec E_b^*$, and $\vec E_c^*$ afterwards in the same order.   
However, there are no simple relations among the various critical times. 
Numerically, one finds $t_{c,1} = 0.7172921525032698574(1), t_{c,2} = 1.25663706143591(1)$, and $t_{c,3} = 2.23933357406560946(1)$.   
Unlike the previous cases, the RG flow starting from the critical times does not seem to go into an unstable fixed-point, but appears to be chaotic, as shown in Fig. \ref{fig:rg_n5}. 

\begin{figure}[hth]
\centering
  \includegraphics[scale=0.40]{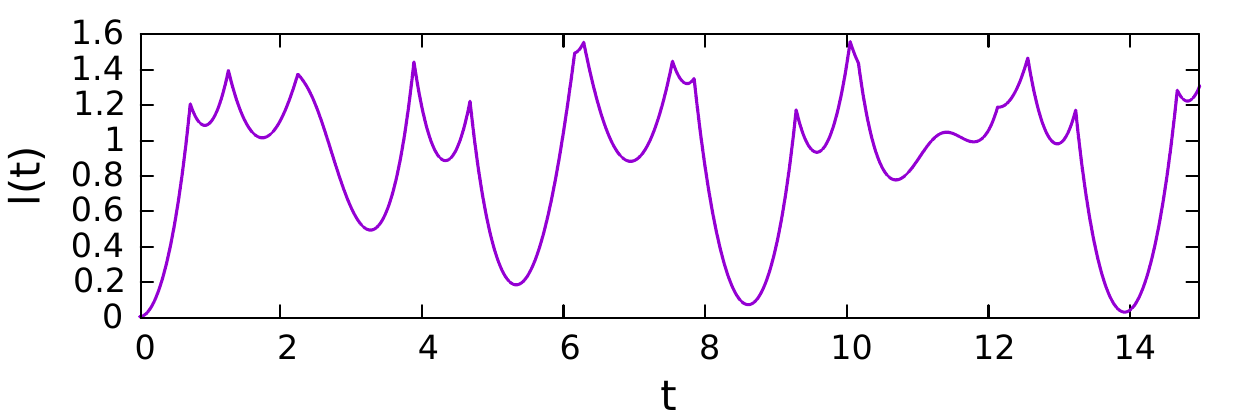}
\caption{The rate function $l(t)$ for $Q = 5$.
}
\label{fig:rate_n5}
\end{figure}

These critical times are confirmed by an exact computation of $l(t)$ by the transfer matrix of $G(t)$.    
The $l(t)$ calculated also appears to be singular at a random sequence of critical times, shown in Fig. \ref{fig:rate_n5}.    
Because there is not an unstable fixed-point which controls the RG flow at the critical times, the value of the critical exponent cannot be obtained straightforwardly. 
However, numerically inspecting the singularity of $l(t)$ in Fig. \ref{fig:rate_n5} shows that $l(t)$ still has a linear cusp near the critical times.     
This can be related to the ``escape time'' of the chaotic RG flow in the following way. 
For $t$ close to a critical time, define the {\it escape time}, $n_e$, of the chaotic part of the RG flow to be the number of RG iterations before the flow eventually settles into the vicinity of the (meta)stable fixed-point.   
For example in the left panel of Fig. \ref{fig:rg_n5}, for $t$ in the left vicinity of $t_{1,c}$, we operationally define $n_e$ as the first RG iteration at which the real part of $E'_1$ exceeds 1.  
In one RG iteration, because the RG transformation preserves the value of the Loschmidt amplitude (up to a regular quantity associated with $(\vec T'_\tmp)_{0s}$) and that the system size decreases by a factor of $b^d$, the singular part of the rate function increases by a factor $b^d$.     
Thus, after $n$ levels of RG iterations, the singular part of the rate function, $l_s(\tau)$, scales as   
\begin{equation}
  l_s(\tau) = b^{-nd}l_s(\vec E^{(n)}) = b^{-n_ed}l_s(\vec E^*)
\end{equation}
where $\vec E^{(n)}$ is the coupling constant after $n$ RG iterations. 
$\vec E^{(n)}$ eventually becomes close to $\vec E^*$, the coupling constant at the (meta)stable fixed-point, after $n_e$ steps. 
Here $n_e$ depends on $\tau$.  
Thus, assuming a power-law singularity of $l(\tau) \sim |\tau|^\alpha$, we obtain   
\begin{equation}
  n_e d = -\log_b \abs{\tau}^\alpha + c = -\frac{\alpha}{\log b} \log\abs{\tau} + c
\end{equation}
Fitting the numerical data for $t$ on the left vicinity of $t_{1,c}$ gives 
\begin{equation}
  n_e = -1.439\log\abs{\tau} + 1.595 
\end{equation}
whereas $1/\log(2) = 1.4427$. 
Despite the crude definition of $n_e$, the two results agree quite well. 
As the singularity of $l(t)$ arises from the level crossing of the dominant and sub-dominant eigenvalues of a finite dimensional transfer matrix, it should generically be a linear cusp.  
Thus, quite remarkably, the above RG analysis serves as a {\it proof} to the relation between $\tau$ and $n_e$ in the chaotic behavior of the recursion relation Eq. \ref{eq:RG}, which would have been difficult to guess.  
\subsection{$Q>5$}
We very briefly sketch the results for $Q > 5$. 
For $Q = 6$, $l(t) = l(t + 2\pi)$, and there are four (meta)stable fixed-points, each of which controls a noncritical phase. 
At the critical times which separate these noncritical phases, the system flows into unstable RG fixed-points whose leading RG eigenvalues are all 2. 
For $Q > 6$, however, the rate function seems to generically have an aperiodic sequence of critical times, starting from which the RG flows are chaotic.  
The aperiodicity of the rate function can be understood from the fact that in the initial coupling constant $E_m = e^{it2\cos(\frac{2\pi}{Q}m)}$, the exponents $\cos(\frac{2\pi}{Q}m)$ are rational for all $m$ only when $Q = 2, 3, 4$ and $6$.     
It, however, remains to be understood why the aperiodicity of the rate function and the chaos of the RG flow occur together.      
We defer this question to future study.  

\section{The pure Potts model}
\label{sec:pure_Potts}
Consider now the $Q$-state Potts model with Hamiltonian 
\begin{equation}
  H_{\Potts} = - \frac{1}{Q} \sum_{i=1}^L \sum_{q = 0}^{Q-1} \sigma_i^q \sigma_{i+1}^{Q-q} - f\sum_{i=1}^L\sum_{q=0}^{Q-1}\tau_i^q
  \label{eq:potts}
\end{equation} 
where $\sigma_i$ and $\tau_i$ are the same as in Eq. \ref{eq:clock}.   
We again take the transverse field, $f$, to be infinite in the pre-quenched Hamiltonian, and zero in the post-quenched Hamiltonian. 
Instead of Eq. \ref{eq:transfer},  the transfer matrix of the Potts model is 
\begin{equation}
  \vec T_{m_im_{i+1}} = E_m = \frac{1}{Q} e^{it \delta_{m 0}}.
  \label{eq:transfer_potts}
\end{equation}
One can always normalize the transfer matrix such that $\T_{00} = 1$. 
Then there is only one independent coupling constant, $x$, in the transfer matrix: 
\begin{equation}
  T_{m_im_{i+1}} = \begin{cases} 1 & m_i = m_{i+1} \\ x & m_i \not= m_{i+1}
  \end{cases} 
  \label{eq:T_potts}
\end{equation}
where $x$ is a complex number. 
As one can check, the renormalized transfer matrix $\vec T'$ by Eq. \ref{eq:RG} still takes the form of Eq. \ref{eq:T_potts}.   
This reduces the renormalization of the transfer matrix to the renormalization of just one coupling constant: \begin{equation}
  x' = \frac{2x + (Q-2) x^2}{1 + (Q - 1)x^2} 
  \label{eq:RGE}, 
\end{equation}
whose starting point is $x^{(0)} = e^{-it}$. 
The fixed point equation of Eq. \ref{eq:RGE}, $x' = x$, has three solutions: 
$x_1 = 0$, $x_2 = 1$, and $x_3 = \frac{1}{1-Q}$.   
The Jacobian of the RG transformation at these three fixed-points are respectively $2$, $0$, and $0$, suggesting that they are respectively unstable, stable, and stable RG fixed-points. 

When $t = 0$, $x$ flows into $x_2 = 1$ for all $Q$.  
Since a DQPT separates different stable dynamical phases of the system, in order for the DQPT to happen, there must be time at which $x$ flows into $x_3 = \frac{1}{1-Q}$. 
However, for infinitely large $Q$, Eq. \ref{eq:RGE} becomes $x' = 1$ regardless the value of $x$, and $x_3$ can never be reached. 
Thus, for sufficiently large $Q$, DQPTs can never occur.
When $Q = 2$, the Potts and the clock model are equivalent, and the DQPT does occur.   
Therefore, there must exist a $Q_c$ for which the DQPT occurs for $Q < Q_c$ and does not occur for $Q > Q_c$.  
Although $Q_c$ does not have to be an integer, it turns out to be exactly 4. 

\begin{figure}[hth]
\centering
\includegraphics[scale=0.4]{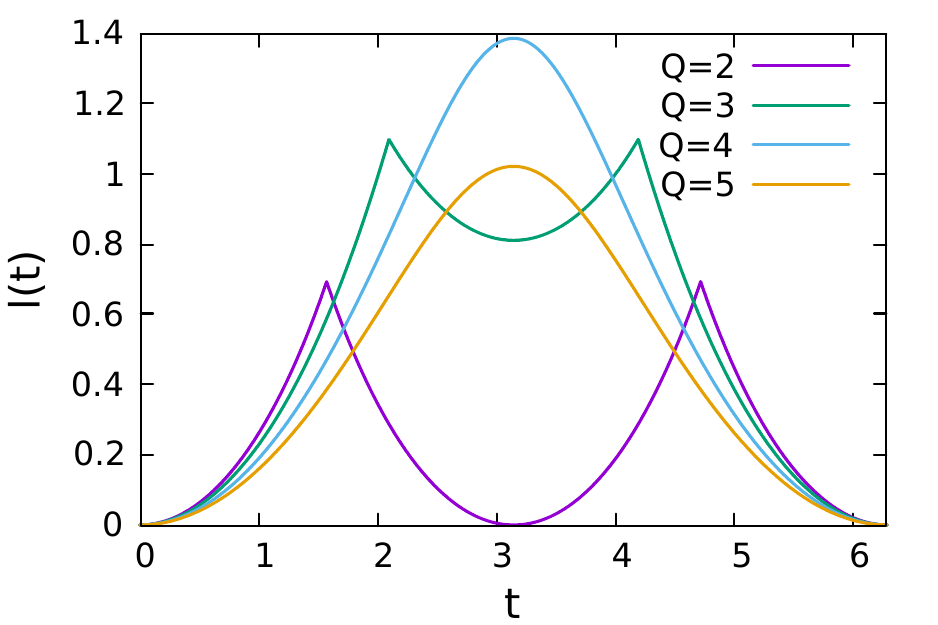}
  \caption{The rate function for the $Q$-state Potts model, obtained through the dominant eigenvalue of the transfer matrix in Eq. \ref{eq:transfer_potts}.
  Note that this transfer matrix is $2\pi$-periodic in $t$. 
}
\label{fig:potts}
\end{figure}

In Fig. \ref{fig:potts}, we show the Loschmidt rate function for $Q = 2, 3, 4$, and 5. 
It is clear that $ 3 < Q_c \le 4$. 
At $t = \pi$, $x^{(0)} = -1$ for all $Q$, and simulating Eq. \ref{eq:RGE}, one discovers that $x$ tends to $\frac{1}{1-Q}$ for $Q < 4$, which would imply a DQPT if $Q$ were integral. 
We thus conclude $Q_c = 4$.
When $Q = 3$, note also that the stable RG fixed-point of the clock model $\vec E_a^* = (1, 1, 1)$ and $\vec E^*_b = (1, -\frac{1}{2}, -\frac{1}{2})$ coincide respectively with $x_2$ and $x_3$. 
Thus, the DQPTs that the Potts chain does experience are identical to the ones in the clock chain. 

\section{The disordered model}
\label{sec:disorder}
In equilibrium critical phenomena, RG analysis often sheds light on the effect of disorder on phase transitions of the pure system. 
In the Harris criterion \cite{Harris, Lubensky}, for example, one checks whether the distribution of random couplings becomes narrower or broader as the coarse-graining iterates at the unstable fixed-point of the pure model, which determines whether disorder is relevant at the phase transition. 
Here we show that the non-equilibrium RG fixed points also control the DQPT of the disordered systems, which are started to be studied only very recently in DQPTs \cite{DQPT_RFIM,DQPT_MBL}.     
Unlike the Harris criterion, as will be shown, it is the stable fixed points that control the DQPT of the disordered system.  
We will modify Eq. \ref{eq:RG} to treat the disordered system. 

Because the DQPTs in the Potts chain are the same as the ones in the clock chain, we study here only the latter.  
Consider now the disordered clock chain where the nearest-neighbor bonds at different lattice sites are drawn independently from a probability distribution, $P(J_i)$.  
We take $\ket{\psi_0} = \ket{\psi}_\product$.
In analogy with the free energy, the self-averaging quantity here should be the quench-averaged rate function: 
\begin{equation}
  \qbraket{l(t)} = \int d\vec J P(\vec J) l_{\vec J}(t). 
\end{equation}
$\vec J = \{J_1, J_2, ...\}$ is one realization of the bonds with a rate function $l_{\vec J}(t)$, and $P(\vec J) = \prod_i P(J_i)$ is the probability density of this realization. 
$\qbraket{\cdot}$ denotes quench-averaging under $P(\vec J)$. 

For concreteness, let us take $Q = 2$ and generalize the results later for other $Q$s. 
When $Q = 2$, the coupling constants can be made all real by coarse-graining the transfer matrix once: 
\begin{equation}
  \begin{split}
    \T^\site{i'} &\propto \begin{pmatrix} e^{i2J_it} & e^{-2J_it} \\ e^{-i2J_it} & e^{2J_it} \end{pmatrix}  \begin{pmatrix} e^{i2J_{i+1}t} & e^{-2J_{i+1}t} \\ e^{-i2J_{i+1}t} & e^{2J_{i+1}t} \end{pmatrix} \\ 
    &= 2 \begin{pmatrix} \cos(2(J_i + J_{i+1})t) & \cos(2(J_i - J_{i+1})t) \\ \cos(2(J_i - J_{i+1})t) & \cos(2(J_i + J_{i+1})t)\end{pmatrix}.
 \end{split}
 \label{eq:real}
\end{equation}
We will thus take $\T^\site{i}$ to be real in the following for notational convenience.   
As seen for $Q = 2$, there are two stable RG fixed points, $\vec E^*_a = (1, 1)$ and $\vec E^*_b = (1, -1)$. 
The attractive basin for $\vec E^*_a$ is $\vec E = (1, a), a > 0$, and for $\vec E^*_b$ is $\vec E = (1, b), b < 0$. 
After step 1 of Eq. \ref{eq:RG} of the coupling constants at two lattice sites, $\vec E^\site{i} = (1, x_i)$ and $\vec E^\site{i+1} = (1, x_{i+1})$, one obtains $\vec E'^{\site{i'}}_\tmp = (1+x_ix_{i+1}, x_i + x_{i+1})$.    
Thus, within the attractive basin of each stable fixed-point, $E'^{\site{i'}}_{0,\tmp} \ge 1$ and the RG equation is perfectly regular.  
In addition, as long as both of $\vec E^\site{i}$ and $\vec E^\site{i+1}$ are in the same attractive basin, their renormalized coupling constant will be closer to the respective stable fixed-point than either $\vec E^\site{i}$ or $\vec E^\site{i+1}$. 
However, when $\vec E^\site{i} = \vec E^*_a$ and $\vec E^\site{i+1} = \vec E^*_b$, step 1 of Eq. \ref{eq:RG} gives, in the form of transfer matrices,  
\begin{equation}
  \begin{pmatrix}1 & 1 \\ 1 & 1\end{pmatrix}
  \begin{pmatrix}1 & -1 \\ -1 & 1\end{pmatrix} 
= 
  \begin{pmatrix}0 & 0 \\ 0 & 0\end{pmatrix}, 
\end{equation}
which makes the second step of Eq. \ref{eq:RG} singular. 
As the RG procedure proceeds, the coupling constants of the disordered chain very quickly settle into the vicinity of one of the two stable fixed-points, and the RG procedure eventually fails.   

To overcome this failure, one first notes that the transfer matrices at different sites commute.   
Consequently, we can move all the transfer matrices in the attractive basin of $\vec E^*_a$ to the left side of the chain, and those in the attractive basin of $\vec E^*_b$ to the right side. 
The $\vec E^*_a$ and $\vec E^*_b$ side of the chain can then be respectively renormalized into one transfer matrix without incurring any singularity:   
\begin{equation}
  \T_a = \begin{pmatrix}1 & 1+\epsilon_a \\ 1 + \epsilon_a & 1\end{pmatrix},   \T_b = \begin{pmatrix}1 & -1+\epsilon_b \\ -1 + \epsilon_b & 1\end{pmatrix}
    \label{eq:Ta}
\end{equation}
where if there are sufficiently many transfer matrices on both sides before the renormalization, $\abs{\epsilon_a} \ll 1$ and $\abs{\epsilon_b} \ll 1$.   
In the process, regular parts of the rate function, $(\T^{\site{i'}}_\tmp)_{0s}$, will be extracted due to step 2 of Eq. \ref{eq:RG}. 
All of the singularity resides in $\T_a$ and $\T_b$.  

To clarify the above RG procedure, we decompose the quench-averaged rate function as follows   
\begin{equation}
  \begin{split}
    \qbraket{l(t)} & = l_0 + \qbraket{l_l(t)}  + \qbraket{l_r(t)} + \qbraket{l_s(t)}   
  \end{split}
  \label{eq:decompose}
\end{equation}
where $l_0 = -\frac{2}{L}\log(Q^L)$, and $\qbraket{l_l(t)}$ and $\qbraket{l_r(t)}$ are the two regular parts extracted from $\qbraket{l(t)}$ by the RG procedure on the two sides of the chain.   
$\qbraket{l_s(t)}$ is the singular part of the rate function and is given by  
\begin{equation}
\begin{split}
  \qbraket{l_s(t)} &= -\frac{2}{L}\qbraket{\Re\{\Log\, \Tr(\T_a(t)\T_b(t))\}} 
\\ 
&= -\frac{2}{L} \qbraket{\log \abs{Q(-\epsilon_a + \epsilon_b + \epsilon_a\epsilon_b)}} 
\end{split}
\end{equation}
Any chain can also be viewed as an assembly of $n$ chains of length $L_0 = \frac{L}{n}$.   
One can independently renormalize these $n$ parts and will end up with a chain composed of transfer matrices $\T_{a,1}, ... , \T_{a,n}$, and $\T_{b,1}, ... , \T_{b,n}$.  
These transfer matrices may be different due to the fluctuation in the realization, but are the same in distribution.  
The final $\epsilon_a$ of the full chain will then be 
\begin{equation}
\begin{split}
  \epsilon_a &= \frac{\text{the off-diagonal element of }(\T_{a,1}...\T_{a,n})}{\text{the diagonal element of }(\T_{a,1}...\T_{a,n})} - 1  
  \\
  &= (\frac{-1}{2})^{n-1}\epsilon_{a,1}...\epsilon_{a,n} + \text{higher-order terms} 
\end{split}, 
\label{eq:epsilon_a}
\end{equation}
where $\epsilon_{a,1},...,\epsilon_{a,n}$ are defined by $\T_{a,1},...,\T_{a,n}$ in the same way as in Eq. \ref{eq:Ta}. 
$\epsilon_b$ can also be similarly written. 
In the thermodynamic limit, $\epsilon_a$ and $\epsilon_b$ both approach zero, and the singular part of the quench-averaged rate function will be    
\begin{equation}
  \begin{split}
    \qbraket{l_s(t)} &= -\lim_{L_0, n\rightarrow \infty} \frac{2}{nL_0} \qbraket{\log(Q\abs{\epsilon_{a,1}...\epsilon_{a,n} - \epsilon_{b,1}...\epsilon_{b,n}})}  
  \\
  &= - \lim_{L_0 \rightarrow \infty} \frac{2}{L_0} \qbraket{\log(\max(\abs{\epsilon_{a,1}}, \abs{\epsilon_{b,1}}))}  
  \\
  &= \lim_{L \rightarrow \infty}  \qbraket{\min(-\frac{2}{L}\log\abs{\epsilon_{a}}, -\frac{2}{L}\log \abs{\epsilon_{b}})} 
\end{split}
\end{equation}
Here we have used the fact that there is no difference between $\epsilon_a$ and $\epsilon_{a,1}$ in the thermodynamic limit.  
As $\epsilon_a$ and $\epsilon_b$ scale exponentially with $L$, as seen in Eq. \ref{eq:epsilon_a}, the above limit exists, and $\qbraket{l_s(t)}$ can finally be written as    
\begin{equation}
  \qbraket{l_s(t)} = \min(l_{a}(t), l_{b}(t)) 
  \label{eq:l_s}
\end{equation}
where 
\begin{equation}
  l_{a/b}(t) \equiv -\lim_{L\rightarrow \infty}\frac{2}{L} \qbraket{\log\abs{\epsilon_{a/b}(t)}}. 
  \label{eq:la}
\end{equation}
In Eq. \ref{eq:l_s}, the order of $\min$ and $\qbraket{\cdot}$ can be swapped, because of the self-averaging property of $l_a(t)$ and $l_b(t)$.  
Now, here is the point: because $l_a(t)$ and $l_b(t)$ are respectively calculated from the renormalization of the system in the same stable phase, they should be smooth functions of $t$, provided that $\epsilon_a$ or $\epsilon_b$ does not become zero.   
$\qbraket{l_s(t)}$ thus generically has a linear singularity when $l_a(t)$ and $l_b(t)$ intersect.  
However, when $\epsilon_a$ and $\epsilon_b$ both become zero, the rate function diverges logarithmically. 

\begin{figure}[hth]
\centering
\begin{minipage}{.25\textwidth}
  \includegraphics[scale=0.4]{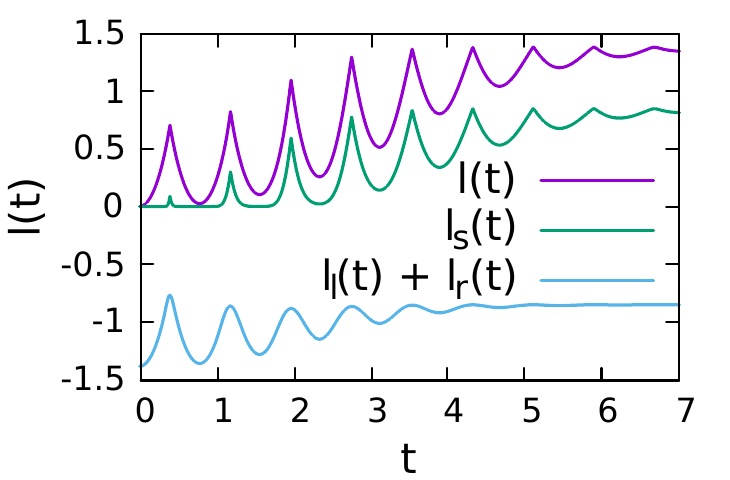}
\end{minipage}%
\begin{minipage}{.25\textwidth}
  \includegraphics[scale=0.4]{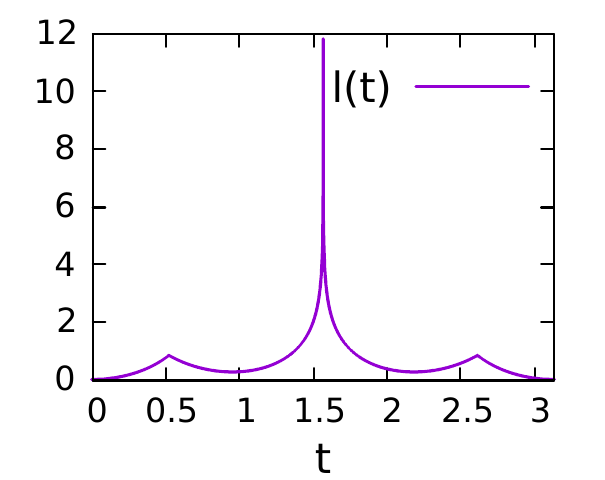}
\end{minipage}%
  \caption{Left: The quench-averaged rate function of the disordered clock model defined by Eq. \ref{eq:Ji}.
The calculation is done for $L = 2^{16}$, and is averaged over $2^{10}$ realizations. 
Right: The rate function of a chain with $J_i = 1$ and $0.5$ each with probability $\frac{1}{2}$.
}
\label{fig:rate_L16}
\end{figure}

\begin{figure}[hth]
\centering
\begin{minipage}{.23\textwidth}
  \includegraphics[scale=0.27]{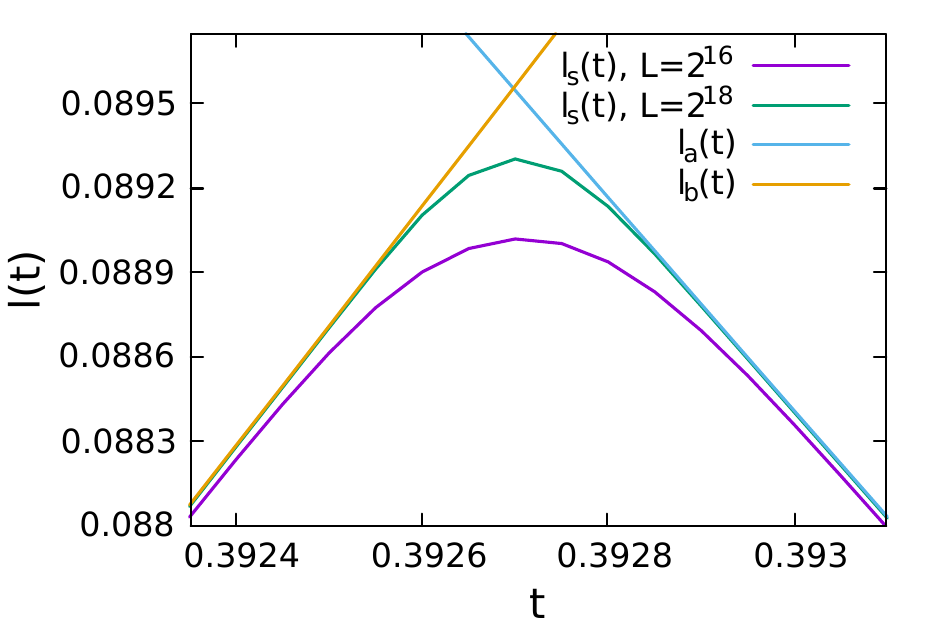}
\end{minipage}%
\begin{minipage}{.23\textwidth}
  \includegraphics[scale=0.27]{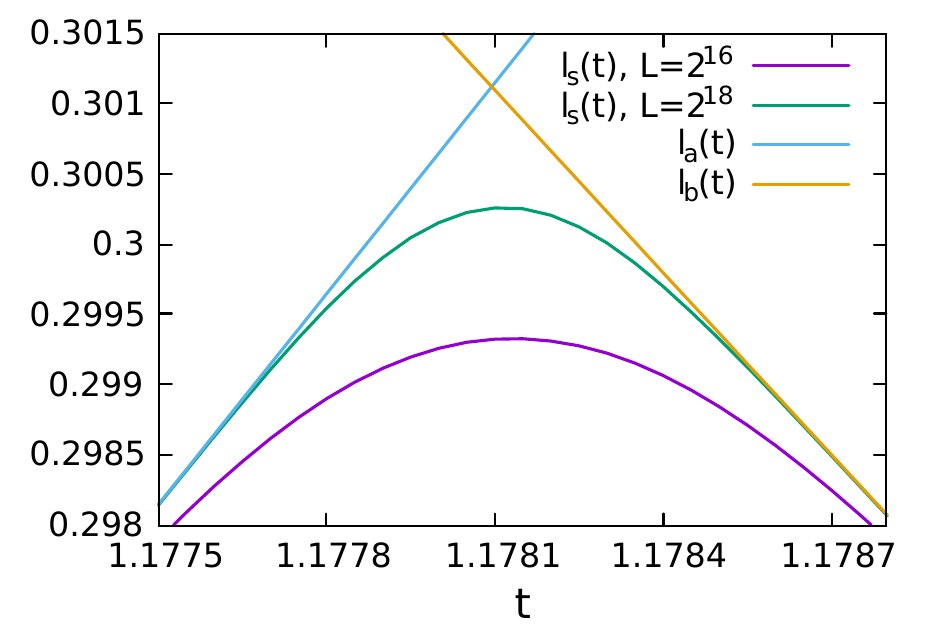}
\end{minipage}%
\caption{$\qbraket{l_s(t)}$, $\qbraket{l_a(t)}$, and $\qbraket{l_b(t)}$ around the first and the second peaks of the $\qbraket{l(t)}$ in Fig. \ref{fig:rate_L16} (left panel).
The $\qbraket{l_s(t)}$ is computed for $2^{16}$ and $2^{18}$ respectively with $2^{14}$ and $2^{13}$ realizations.  
The $\qbraket{l_a(t)}$ and $\qbraket{l_b(t)}$ have very weak size dependences and are only shown here for $L = 2^{18}$. 
}
\label{fig:rate_eps}
\end{figure}

Consider first random bonds of a chain given by  
\begin{equation}
  J_i = 1 + 0.1 g, \hspace{5mm} g \sim \mathcal{Q}(0, 1) 
  \label{eq:Ji}
\end{equation}
independently at each site $i$.
Here $g$ is a unit Gaussian random variable. 
For any realization of the bonds, the various terms of the rate function in Eq. \ref{eq:decompose} can be numerically calculated by the RG procedure.  
An arbitrary precision arithmetic package, such as TTMath \cite{ttmath}, which we use, will be necessary for the calculation of a long chain. 
The result of the calculation is presented in Fig. \ref{fig:rate_L16} (left) and Fig. \ref{fig:rate_eps}.  

If the chain is composed of transfer matrices which renormalize into $\vec E_a^*$ and $\vec E_b^*$ in finite RG iterations, then both $\epsilon_a(t)$ and $\epsilon_b(t)$ become zero.  
For example, when $Q = 2$, consider a chain with $J_i = 1$ and $\frac{1}{2}$ with probability $p$ and $1-p$.   
At $t = \frac{\pi}{2}$, $J_i = 1$ and $\frac{1}{2}$ respectively give coupling constants $\vec E_l = (-1, -1)$ and $\vec E_r = (i , -i)$.   
Under just one iteration of the RG procedure in Eq. \ref{eq:RG}, $\vec E_l$ goes into $\vec E_a^*$ and $\vec E_r$ goes into $\vec E_b^*$.  
This means that $\epsilon_a(t)$ and $\epsilon_b(t)$ are both strictly zero at $t_c = \frac{\pi}{2}$. 
Thus, for $t$ in the vicinity of $t_c$, the rate function is  
\begin{equation}
  l_s(t) \propto -\log(\abs{t - t_c}), \hspace{5mm} \text{   for $t$ close to $t_c$.} 
\end{equation}
This is shown in Fig. \ref{fig:rate_L16} (right).      

We now generalize the result to other $Q$s. 
First note that the commutativity of the clock model transfer matrices still holds for $Q > 2$. 
In addition, at least for $Q = 3, 4$, and $5$, multiplying the stable RG fixed-point transfer matrices with one another gives the zero matrix.  
For example, when $Q = 3$, there are two stable RG fixed-points, $\vec E_a^* = (1,1,1)$ and $\vec E_b^* = (1, -\frac{1}{2}, -\frac{1}{2})$, corresponding to two fixed-point transfer matrices, $\T_a^*$ and $\T_b^*$.
As one can check, $\T_a^* \T_b^* = 0$. 
Then, the arguments from Eq. \ref{eq:Ta} to Eq. \ref{eq:la} follow identically, giving $\qbraket{l_s(t)} = \min(l_a(t), l_b(t))$, where $l_a(t)$ and $l_b(t)$ are analogously defined as in Eq. \ref{eq:la}. 
\label{sec:excited}
\section{Conclusion}
\label{sec:conclude}

In this paper, we studied the renormalization of the transfer matrices of the Loschmidt amplitude of the clock model and the Potts model.  
The fixed-points of this RG procedure are found to determine the DQPT of both the pure and the disordered system.  
Many problems can be investigated in the future under the RG framework established.  
For example, a universality class of DQPT with critical exponent $\frac{1}{2}$ is identified using the current RG procedure \cite{DQPT_Critical}. 
Another question that is most interesting concerns with the dynamical quantum critical region described in \cite{Heyl_2019, DQPT_symmetry_breaking, experiment1}, for which the RG procedure seems a particularly useful tool. 
\begin{acknowledgments}
The author is grateful to Ling Wang for hosting him at the Beijing Computational Science Research Center, introducing him to DQPTs, and many stimulating discussions.  
He is also grateful for mentoring from his advisor Roberto Car at Princeton. 
The author acknowledges support from the DOE Award DE-SC0017865. 
\end{acknowledgments}
\bibliography{abc}
\end{document}